\documentclass[fleqn,usenatbib]{mnras}

\usepackage{newtxtext,newtxmath}

\usepackage[T1]{fontenc}
\usepackage[caption=false]{subfig}

\DeclareRobustCommand{\VAN}[3]{#2}
\let\VANthebibliography\thebibliography
\def\thebibliography{\DeclareRobustCommand{\VAN}[3]{##3}\VANthebibliography}

\usepackage{graphicx}	
\usepackage{amsmath}	

\newcommand{\tconv}{t_{\rm conv}}
\newcommand{\tinsp}{t_{\rm insp}}
\newcommand{\Ebind}{E_{\rm bind}}
\newcommand{\Eorb}{\Delta E_{\rm orb}}

\newcommand{\rconv}{r_{\rm conv}}

\title[High-Mass CE]{Convection Reconciles the Difference in Efficiencies Between Low-Mass and High-Mass Common Envelopes}

\author[Wilson \& Nordhaus]{
E. C. Wilson,$^{1,2}$\thanks{E-mail: wilsone@lycoming.edu}
J. Nordhaus,$^{1,3}$
\\
$^{1}$Center for Computational Relativity and Gravitation, Rochester Institute of Technology, NY 14623, USA\\
$^{2}$Lycoming College, Williamsport, PA 17701, USA\\
$^{3}$National Technical Institute for the Deaf, Rochester Institute of Technology, NY 14623, USA\\}

\date{Accepted XXX. Received YYY; in original form ZZZ}

\pubyear{2022}

\begin{document}
\label{firstpage}
\pagerange{\pageref{firstpage}--\pageref{lastpage}}
\maketitle

\begin{abstract}
The formation pathways for gravitational-wave merger sources are predicted to include common envelope (CE) evolution. Observations of high-mass post-common envelope binaries suggest that energy transfer to the envelope during the CE phase must be highly efficient.  In contrast, observations of low-mass post-CE binaries indicate energy transfer during the CE phase must be highly inefficient.  Convection, a process present in low-mass and high-mass stars, naturally explains this dichotomy.  Using observations of Wolf-Rayet binaries, we study the effects of convection and radiative losses on the predicted final separations of high-mass common envelopes.  Despite robust convection in massive stars, the effect is minimal as the orbit decays well before convection can transport the liberated orbital energy to the surface.  In low-mass systems, convective transport occurs faster then the orbit decays, allowing the system to radiatively cool thereby lowering the efficiency.  The inclusion of convection reproduces observations of low-mass and high-mass binaries and remains a necessary ingredient for determining outcomes of common envelopes.

\end{abstract}




\section{Introduction}

When binary star systems evolve, the more massive star's ascent up the giant branches may initiate a common envelope (CE) phase if the initial separation is $\lesssim 5-10$ AU \citep{Paczynski1976}. Mass loss from the primary star via winds and an expanding radius can disrupt the stability of the orbit, causing the companion to enter the primary's envelope, either through Roche Lobe overflow, orbital decay via tidal dissipation, or direct engulfment \citep{Nordhaus2010, Nordhaus2013, Ivanova2013, Kochanek2014, Chen2017}. Immersed within a shared, common envelope, angular momentum and energy transfer from the orbit to the envelope, resulting in a rapid reduction in the separation between the companion and primary's core.  The final outcome of a CE phase is either (i.) ejection of the envelope and the emergence of a tight binary \citep{Iben1993,Nordhaus2006,Nordhaus2007}, or (ii.) destruction of the companion (or core) and the emergence of a single star whose evolution has been significantly altered by the interaction \citep{Nordhaus2011,Guidarelli2019,2021arXiv210500077G}.

While other mechanisms, such as the Kozai mechanism, tidal friction, \citep{Fabrycky2007,Thompson2011,Shappee2013} and mass-loss from a third body, may produce tight binaries in triple and higher-order systems \citep{Michaely2016}, CE evolution is thought to be the primary formation-channel by which short-period (typically $\lesssim$ 1 day) binaries are produced \citep{Toonen2013,Canals2018,Kruckow2018}. Despite its importance, the CE phase is poorly understood. Observations of systems entering, undergoing, and exiting the CE phase are challenging as the orbital decay timescales are short (months to years), the rate in the galaxy is low ($\sim$1 per decade), and the predicted transient signatures are faint, reducing the likelihood of direct detection \citep{2017ApJ...834..107B,Jones2020}.  Instead, observations of precursor emission \citep{MacLeod2018} and post-CE phenomena such as short-period binaries \citep{Podsiadlowski1992,yoon2015}, compact-object binaries \citep{Vigna-Gomez2018}, X-ray binaries \citep{Kalogera1998,VanDenHeuvel,Tauris2017}, Wolf-Rayet binaries, planetary nebulae, stripped-envelope supernovae, gravitational-wave sources \citep{Mandel2021}, and other systems are used to indirectly study CEs \citep{Ivanova2013}.

In addition to observational studies, numerical simulations of CEs are an active field of research.  Early simulations of a massive star primary (16 $M_\odot$) with a solar-mass neutron star companion were performed in spherical symmetry \citep{Taam1978}, axisymmetry \citep{Bodenheimer1984}, and in three dimensions \citep{Terman1995}. Recent simulations of massive star CEs have focused on ejection mechanisms such as radiation pressure \citep{Fragos2019,Lau2022}, winds, recombination energy \citep{Ricker,Moreno2021}, and jets \citep{Hillel2021} in the envelope. 

While the change in orbital energy is often sufficient to unbind the envelope in both low-mass and high-mass CEs, many numerical simulations find that the envelope remains bound, or is only partially ejected when only orbital energy is considered \citep{Ricker2012,Passy2012,Ohlmann2015,Chamandy2018}.  This has led to numerous studies that investigate tapping additional energy sources to power ejection, i.e. from accretion/jets or recombination, as well as from longer-term processes \citep{Ivanova2015,Soker2015,Kuruwita2016,Sabach2017,Glanz2018,Grichener2018,Ivanova2018,Kashi2018,Soker2018,Reichardt2020,Schreier2021}. 

It should be noted that, due to computational complexity, current simulations neglect prominent physical processes such as convection and radiation.  High-mass and low-mass (super)giant stars possess deep and vigorous convective envelopes capable of redistributing the orbital energy and in many cases transporting it to the surface where it can be radiated away.  When this occurs, the CE can radiatively cool, allowing orbital decay to continue until convective transport to the surface is no longer feasible.  Including convection in CE modelling has been shown to reproduce the observed post-CE populations of M-dwarf+white dwarf binaries and double white dwarf (DWD) systems \citep{Wilson2019,Wilson2020}.

In this paper, we investigate the effects of convective transport on massive star common envelope evolution.  Using detailed stellar interior models, we compare the orbital decay timescales to the convective transport timescales for the observed population of Wolf-Rayet (WR) binaries. In contrast to low-mass stars, massive star CEs are highly efficient as the orbital decay timescale is often shorter then the time required for convection to transport the energy to the surface. When convection is included, we recover the observed Wolf-Rayet orbital separations and provide an explanation for the difference in efficiencies observed in low-mass and high-mass CEs. 

In Section~\ref{sec:Motivation}, we describe an analytical approach to determining the effects of convection on low-mass CE outcomes. In Section~\ref{sec:Methods}, our methods for investigating convection in massive-star CEs are presented. We discuss our results in Section~\ref{sec:Results} and conclude in Section~\ref{sec:Conclusion}.

\section{Motivation}
\label{sec:Motivation}
\subsection{Energy Efficiency in CEs}

Whether a common envelope phase results in the emergence of a short-period binary is often estimated with widely-used energy arguments that compare the binding energy of the envelope to the energy liberated as the orbit decays. During inspiral, orbital energy is transferred to the envelope and, if not lost from the system, must contribute to unbinding.  If sufficient energy is supplied, the envelope is ejected and a compact, post-CE binary emerges.

The efficiency with which orbital energy can be used to drive ejection, $\alpha$, is typically defined as
\begin{equation}
	\alpha=E_{\rm bind}/\Delta E_{\rm orb},
\end{equation}
where $\Ebind$ is the binding energy of the shared envelope, and $\Eorb$ is the change in orbital energy \citep[e.g.,][]{Webbink1984,DeMarco2011}. Binary population synthesis studies, which typically take $\alpha$ as a constant, indicate the efficiency of the CE interaction of low-mass binaries must be low to reproduce observations \citep{Zorotovic2010,Toonen2017}. However, even with low and constant values for the efficiency, population studies over-produce long-period binaries in contrast to observations \citep{Davis2010}. 

Similar studies of high-mass binaries also employ $\alpha$ to describe the ejection efficiency of more-massive systems. In both simulations and population synthesis studies, the CE phase in high-mass stars is usually found to be highly efficient \citep[e.g.,][]{Tauris1996a,Fryer1999,Moreno2021,Fragos2019,Vigna-Gomez2022} with few exceptions \citep{Law-Smith2020}. 

Recently, there is mounting evidence suggesting that the efficiency is not constant and depends on properties of the binary and the internal structure of the envelope \citep{Iaconi2019,Wilson2019,Wilson2020}.  For example, giant stars possess substantial convective envelopes that are capable of transporting and mixing energy from deep in the interior of the star to the surface.  How, and where, energy is released during CE evolution, and how it is transported, can have profound implications on the outcomes.  If convection can rapidly carry the energy that is released as the orbit decays to the surface, it can be radiated away and have little effect on ejection.  However, if convection cannot transport the orbital energy, it remains trapped in optically thick gas and must be used to aid ejection.  Understanding where, and how, the energy of the CE interaction is released and redistributed, particularly recombination energy \citep{Ivanova2015,Sabach2017,Grichener2018,Ivanova2018,Soker2018} and orbital energy \citep{Wilson2019,Wilson2020}, is an area of active research.

\subsection{The Effect of Convection on Low-Mass Common Envelope Evolution}
\label{sec:PapersIandII}

Motivated by observations of low-mass post-CE systems that demonstrate the inefficiency of the common envelope phase, we previously studied how energy is carried by convection in low-mass common envelopes \citep{Wilson2019,Wilson2020}.  Here, we apply similar techniques to study the effect of convective transport on the predicted final separations for massive star CEs.  We begin by considering the amount of energy per unit time released by the companion during its inspiral through the envelope:
\begin{equation}
    L_{\rm drag}=\xi \pi \rho R_{\rm acc}^2(v-v_{\rm env})^3,
    \label{eq:Ldrag}
\end{equation}
where $R_{\rm acc}$ is the accretion radius, given by $R_{\rm acc}=\frac{2 G m_{\rm comp}}{(v-v_{\rm env})^2+c_s^2}$ \citep{Nordhaus2006}. The radially dependent density of the star is given by $\rho$ and the term $\xi$ is a dimensionless factor that depends on the Mach number of the companion's motion through the envelope \citep{1985MNRAS.217..367S}. Because the motion is subsonic everywhere, we assume $\xi=4$. The velocity terms $v_r$, $c_s$, and $v_\phi$ are the radial velocity of the companion, the sound speed, and the Keplerian velocity, respectively, such that $v=(v_r, v_\phi, 0)$. We assume $v_r\ll v_\phi$ at all times (i.e., $v \approx v_\phi$). 

As a limiting case, we assume that the envelope is stationary at the onset of the CE phase ($v_{\rm env}=0$). A rotating envelope would decrease the drag luminosity, increase the time it takes the companion to migrate during the CE, and therefore make it easier for convection to remove energy from the system.

The time rate of change of the gravitational potential energy of the binary is the source of the drag luminosity as the companion inspirals through the envelope and is given by:
\begin{equation}
    \frac{{\rm d}U}{{\rm d}t}=\frac{G m_{\rm comp} v_r}{r}\left(\frac{{\rm d}M}{{\rm d}r}-\frac{M}{r}\right),
    \label{eq:dUdt}
\end{equation}
where $M$ is the mass enclosed at radius $r$. Therefore the radial velocity of the decaying orbit can be found by equating Eqn.~\ref{eq:Ldrag} and Eqn.~\ref{eq:dUdt}. This yields
\begin{equation}
    v_r=\frac{4 \xi \pi \rho G m_{\rm comp} v_\phi^3 r}{\left(\left(\rm{d}M/\rm{d}r\right)-\left(M/r\right)\right)\left(v_\phi^2+c_s^2\right)^2}
\end{equation}
which we invert and integrate from the surface of the star, $r=R_\star$, to a point within the envelope, $r=r'$.  The inspiral timescale represents the amount of time necessary for the companion's orbit to decay through the common envelope and is given by:

\begin{equation}
	t_{\mathrm{inspiral}}[r'] = \int^{r'}_{R_\star}{\frac{\left(\frac{\mathrm{d}M}{\mathrm{d}r}-\frac{M}{r}\right)\ 	(v_\phi^2+c_s^2)^2}{4 \xi \pi G m_{\rm comp} r \rho v_\phi^3}\mathrm{d}r }.
	\label{eq:tinsp}
\end{equation}

The inspiral timescale is compared to the convective transport timescale, $\tconv$, defined as the time required for a convective eddy to travel from its current position to the surface of the star.  This timescale is given by
\begin{equation}
	t_{\mathrm{conv}}[r'] = \int_{r'}^{R_{\star}}\frac{1}{v_{\mathrm{conv}}} \mathrm{d}r,
	\label{eq:tconv}
\end{equation}
where $v_{\rm conv}$ is the convective velocity and $R_\star$ is the radius of the stellar surface.

When $\tconv<\tinsp$, convective eddies rise to the surface faster than the orbit decays, allowing the system to self-regulate via radiative cooling.  In these regions, we assume all of the orbital energy is lost, and thus $\alpha=0$. Conversely, where $\tinsp<\tconv$, convective transport is slow compared to inspiral.  As the orbit decays, the liberated energy cannot escape the optically thick envelope and must contribute to ejection.  In these regions, convection acts to distribute the energy throughout the volume and thus we assume $\alpha=1$. Fig.~\ref{fig:timescales} shows an example where the convective transport timescale, is greater than the inspiral timescale through the majority of the envelope, indicating that convection is not effective at transporting the companion's orbital energy.

For the orbital energy to be fully radiated also likely requires that convection retain its subsonic nature.  The maximum amount of luminosity that convection can accommodate before transitioning to the supersonic regime is given by 
\begin{equation}
    L_{\rm conv, max}= 4 \pi \rho r^2 c_s^3
    \end{equation}
\citep{Quataert2012, Shiode2014, Sabach2017}.  In order to accommodate the additional energy, this must be greater than the luminosity released during orbital decay, or the drag luminosity described in Eqn.~\ref{eq:Ldrag}. Where $L_{\rm drag}\le L_{\rm conv, max}$, convection is able to to remain subsonic despite the increased energy transport. In cases where $L_{\rm drag}>L_{\rm conv}$, convection transitions to the supersonic regime whereby internal shocks could contribute to raising the negative binding energy of the envelope.

\section{Methods}
\label{sec:Methods}

Using evolutionary tracks of massive stars ranging between $15M_\odot$ and $70M_\odot$, we calculate the final predicted separations of post-CE binaries when convection and radiative losses are included in the CE phase. We compare these predictions to observations of Wolf-Rayet binary candidates where at least one component, but not necessarily both components, is a WR star.

\subsection{Observations}
Our observational sample consists of WR binary candidates in the Small Magellanic Cloud and the Large Magellanic Cloud \citep{shenar2016,Shenar2020}.  In particular, WR binary candidates in the LMC are nitrogen-sequence WR stars (WN).  From this sample, we select the 19 binaries that have measured orbital parameters.  In each of these studies, the parameters of the systems were derived using binary evolution models, based on observations of radial velocities, composite spectra, and x-ray luminosity data. Each system has a WR primary and companions of various types including main-sequence stars, giants, or WR stars, determined via luminosity class and binary evolution modelling.  Seventeen of our primary WR stars have masses between 9 and 66$M_\odot$, one has a mass of 2 $M_\odot$, and one has a mass of 139 $M_\odot$.  Most systems have periods between ${\sim}$2 and ${\sim}$ 34 days.  Four systems have periods $>92$ days.

\subsection{Stellar Models}

We produce spherically-symmetric models of massive star interiors with the open-source, stellar evolution code Modules for Experiments in Stellar Astrophysics \citep[\texttt{MESA}, release 11701,][]{Paxton2010,Paxton2018,Paxton2019}. The full evolution of high-mass stars with solar metallicity ($Z=0.02$) and the following zero-age-main-sequence (ZAMS) masses are calculated: 15, 20, 25, 35, 40, 50, 55, 60, and 70$M_\odot$.  For each model, we extract the radially-dependent internal structure of the star at various epochs in its evolution. The radii at which the profiles were extracted are shown in colour in Fig.~\ref{fig:rvst}(a). 

As a star evolves, the core becomes increasingly dense causing the central binding energy to increase. We limit consideration to models of primary stars which have evolved to have distinct central cores. Similarly, we exclude models which occur after the maximum radius has been achieved.   After the star reaches its maximal extent, the radius contracts, making it likely that the CE phase would have started during a prior stage of evolution.

\begin{figure}
    \centering
    \includegraphics[width=\columnwidth]{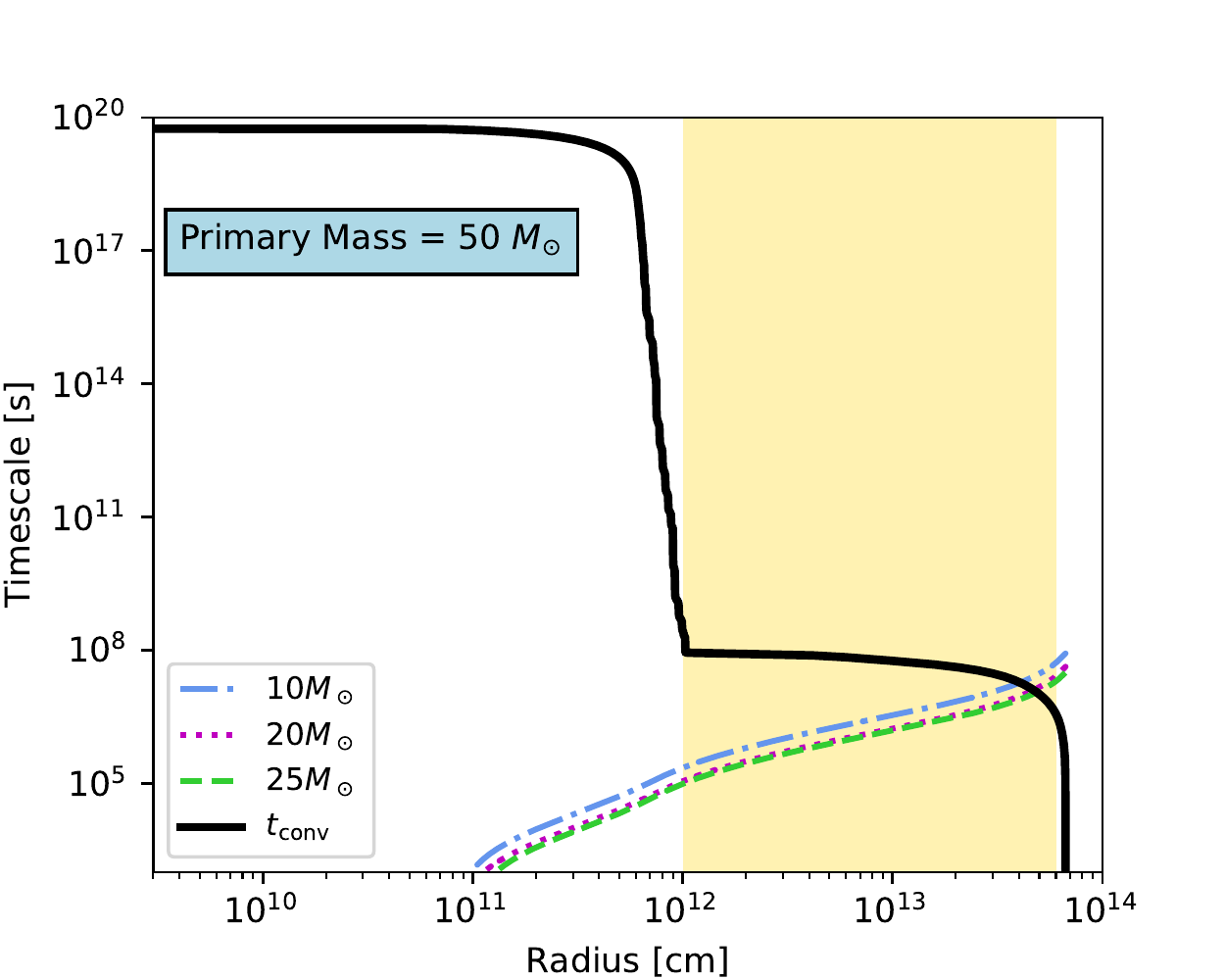}
    \caption{Convective transport timescale (thick, black curve) and inspiral timescales (dashed, coloured curves) of a $50 M_\odot$ primary and three companions. Convective zone is shaded yellow. Convection is only effective at carrying energy to the surface of the envelope where $t_{\rm conv}<t_{\rm insp}$. The efficiency in this example is high as the inspiral timescales are short compared to the convective transport timescale throughout nearly the entire envelope. }
    \label{fig:timescales}
\end{figure}

Following the development of a core, the convective regions of high-mass stars are determined for each point in the primary's subsequent evolution (see Fig.~\ref{fig:rvst}(a)). In the envelopes of massive stars, and occasionally deep into the envelope, convection is well developed and vigorous.  To determine where convection can be effective at transporting the liberated orbital energy of the companion's decaying orbit to the surface of the star, the convective transport time (Equation \ref{eq:tconv}) is compared to the inspiral time (Equation \ref{eq:tinsp}) for various companion masses. Where $\tconv<\tinsp$, the convective transport timescale dominates over the inspiral timescale and convection can quickly carry energy to the surface of the primary where it can be radiated away. Convective eddies can effectively transport a maximum amount of energy before the convection becomes supersonic. This limit is described in Section~\ref{sec:PapersIandII}. At radii where $L_{\rm drag}>L_{\rm conv, max}$ and/or $\tinsp<\tconv$, $\alpha=1$ as convection cannot effectively carry all energy in these regions. In Fig.~\ref{fig:timescales}, the convective transport timescale as a function of radius is compared to several inspiral timescales. In these examples, the inspiral times are shorter than the convective transport times throughout the majority of the envelope, thereby demonstrating that convection is unable to carry energy deposited by the companion away from the system.

\subsection{Final Separations in a Dynamic Envelope}
\label{sec:finalsepshift}

Companions with masses ranging from 10$M_\odot$ to 35$M_\odot$ (in increments of $5 M_\odot$) were allowed to be engulfed in modelled primary stars of various masses (15, 20, 25, 35, 40, 50, 55, 60, and 70$M_\odot$) at different times in the evolution. We define the envelope ejection radius, $r_{\rm ej}$, as the position in the CE where the energy deposited by the companion's shrinking orbit has exceeded the energy required to unbind the envelope.  Formally, this is the maximum separation, $r$, inside the CE where $\alpha\Eorb[r] \ge \Ebind[r]$. If $r_{\rm ej}$ lies within the convective region of the star, the envelope begins to unbind but stellar material interior to the initial ejection radius expands outwards \citep{Soberman1997}. In this case, the final separation of the companion and WR star, $a_{\rm final}$, is determined to be the radius at which the orbital energy of the companion exceeds the binding energy at the boundary between the radiative and convective layers.  This occurs where $\alpha \Eorb[a_{\rm final}]=\Ebind[r_{\rm conv}]$.  Here, $\rconv$ denotes the radius of the boundary between the radiative and convective zones in the primary, and $a_{\rm final}$ is always greater than the radius of the compact core. Should this $a_{\rm final}$ be interior to the radiative zone it is redetermined to be $\rconv$ because ejection-expansion feedback ceases in a radiative region. Companions that tidally shred prior to providing sufficient energy to eject the envelope are excluded because they do not form a post-CE binary.

\section{Results \& Discussion}
\label{sec:Results}

\begin{figure}

\subfloat[The radial evolution of several high-mass stars is shown. The colour of the curve corresponds to the radius of the stellar evolution model. While colour and radius are redundant for individual curves on this plot, the colours shown here directly correspond with the colours shown in (b).]{\includegraphics[clip,width=\columnwidth]{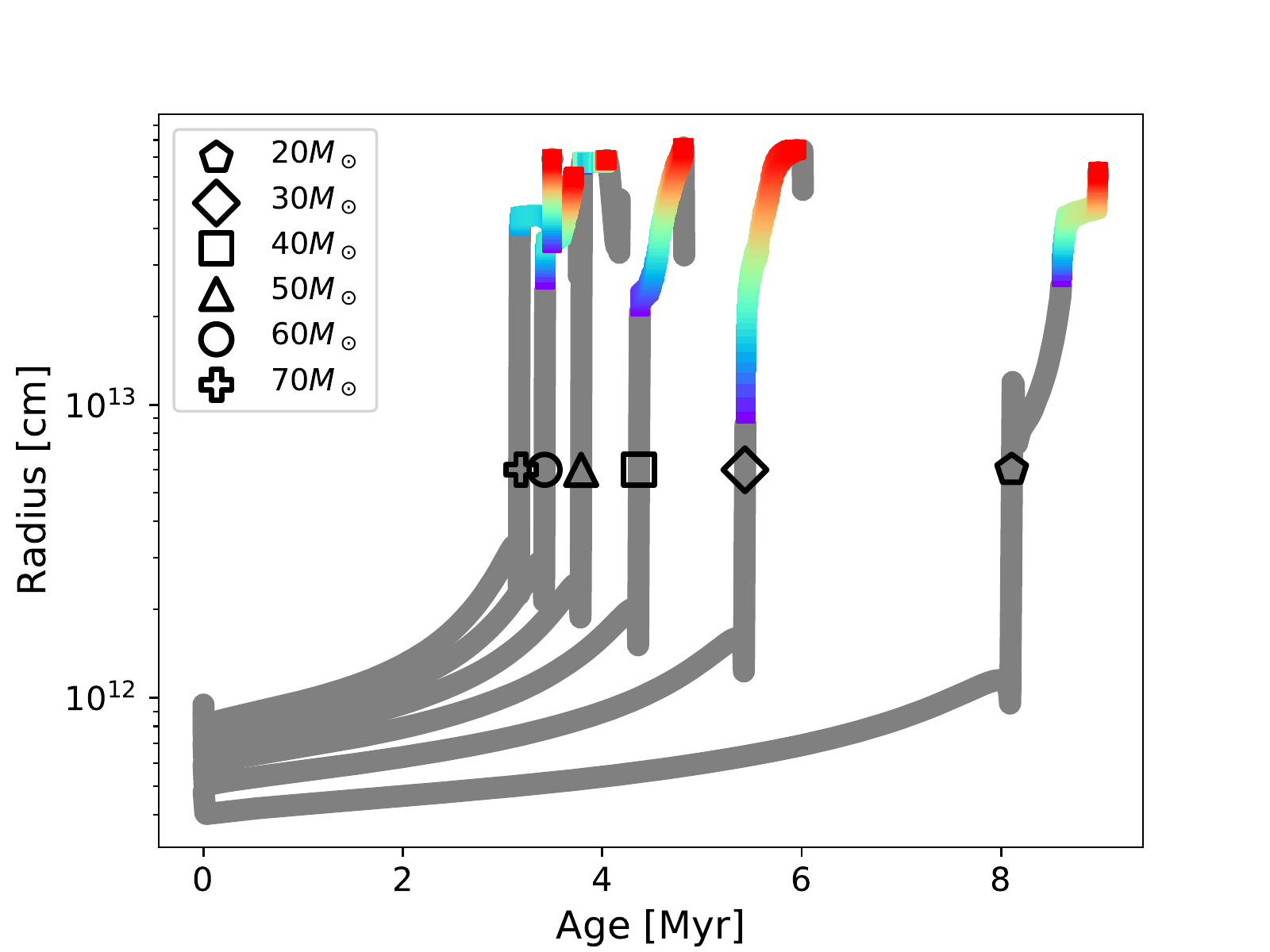}}

\subfloat[Orbital separations of observed WR+MS are shown in black crosses and WR+giant stars (i.e., luminosity classes I-III) are shown in white crosses. The ZAMS mass of the modelled primary is indicated by the symbol shape while the final Wolf-Rayet mass is shown on the x-axis. The final orbital separation between the modelled primary and companion is determined while incorporating effects of convection. ]{\includegraphics[clip,width=\columnwidth]{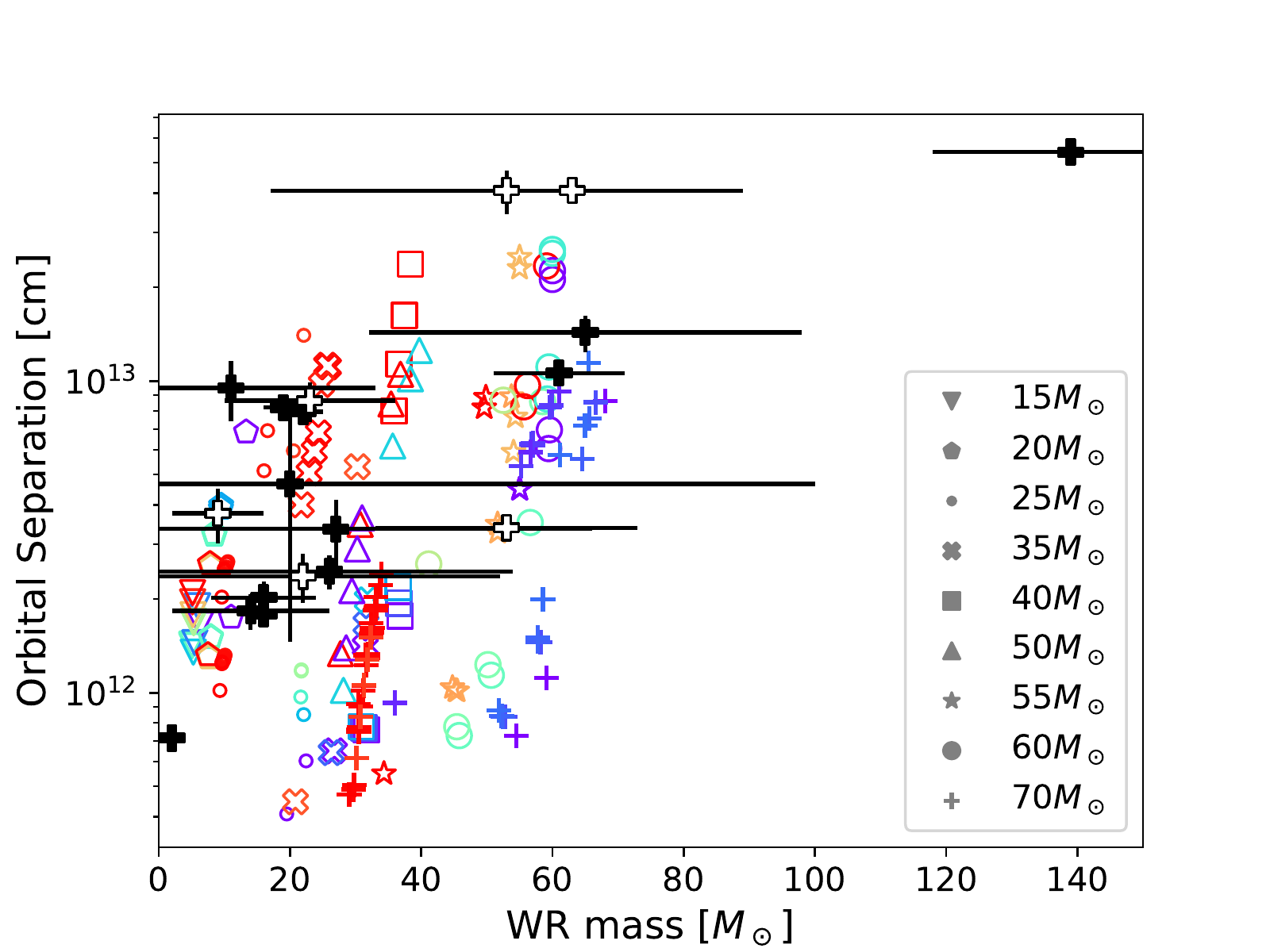}}
    \label{fig:sepvsmass}
\caption{The colour of the modelled points in (b) corresponds to the colour shown in (a), such that the radius of the primary at the time of engulfment can be known.}
   \label{fig:rvst}
\end{figure}

Given the final orbital separation, we determine the mass of the stripped star to be the mass enclosed at the final separation, or $M[a_{\rm final}]$. Our predictions for post-CE binaries consisting of surviving MS companions around stripped-envelope primaries were then compared to observations of Wolf-Rayet binaries. We derive the orbital separations for observed systems assuming circular orbits. The majority of the observed WR binary systems are matched (within error) by at least one of our theoretical post-CE systems.  This comparison is shown in Fig.~\ref{fig:rvst}(b). Initial primary masses (ZAMS masses) are indicated with different symbols while the final mass of the stripped star, $M_{WR}$, is displayed on the horizontal axis. The orbital separation, $a$, between the companion and the WR star is presented on the vertical axis.  Symbol colours indicate the radius of the primary at the time of engulfment as indicated in Fig.~\ref{fig:rvst}(a). If the presence of a companion were to maximize the speed of convection ($v_{\rm conv}=c_{\rm s}$), thereby lowering the convective transport timescale, the difference between the presented final orbital separations and the  final orbital separations when $v_{\rm conv}=c_s$ would be negligible.

The primary masses used to determine the final separations range from 15$M_\odot$ to 70$M_\odot$.  We restricted our analysis to companions such that the mass ratio, $m_2/M_1$, remained less than $0.7$.  Some visual gaps in Fig.~\ref{fig:rvst}(b) may be closed with the use of additional primary mass models.  Note that we did not attempt to model the highest mass observed system. 

While the final separations and masses of individual systems are discernible in Fig.~\ref{fig:rvst}(b), the specific initial conditions which formed the observed binaries remain unknown. For this reason, we analyze the population of modelled systems as a whole.

\subsection{Comparing Low-Mass and High-Mass Results}
 
In two previous studies, we examined the role of convection on two populations of low-mass post-CE systems, white-dwarf+main-sequence (WDMS) binaries and double white dwarfs. Similar to high-mass stars, the envelopes of low-mass (less than 8$M_\odot$) post-main-sequence stars, Red Giant Branch (RGB) and Asymptotic Giant Branch (AGB) stars, are almost fully convective. Generally, when convection is present in low-mass CEs, it significantly lowers the efficiency of the interaction such that companions travel deeper into the envelope, ending in short-period binaries consistent with observations. 

Final separations of systems that undergo a convective common envelope phase are generally smaller than final separations of systems which do not. Radiative losses via convective transport force the companion to travel deeper into the envelope before the CE is ejected, rather than eject the envelope closer to the surface. The final separations of low-mass, post-CE binaries following these convective arguments tend to produce sub-day orbital periods, consistent with observations of M-dwarf+WD binaries \citep{Wilson2019}. In these systems, convective transport to the surface occurs faster than the orbit decays and is able to accommodate the orbital energy deposited by the companion.

When WD companions are allowed to inspiral through a convective CE culminating in formation of a double white dwarf (DWD), similar timescale arguments are relevant. The inspiral timescales for WD companions are shorter than the inspiral timescales for M-dwarf companions, while $L_{\rm drag}$ for WD companions is larger.  For convection to effectively transport the companion's change in orbital energy to the surface: $\tconv<\tinsp$ and $L_{\rm drag}<L_{\rm conv, max}$. In the case of more-compact and massive WD companions, angular momentum transport may spin the envelope up to some fraction of co-rotation.  This acts to increase the inspiral time for the companion making it easier for convection to dominate. WD companions are thus able to travel deeper into the star before ejecting the envelope. When these theoretical final predicted separations are compared to observations of DWD candidates, the parameter space is matched, indicating that convection and radiation play a role in the formation of short-period DWDs \citep{Wilson2020}.

In the high-mass primary cases presented here, the companions' inspiral timescales are typically short compared to the convective transport timescales. This means that the companions contribute sufficient energy to eject the envelope close to the surface of the shared envelope, rather than deep within it, resulting in wider final separations than are seen in the low-mass cases. This is consistent with observations of WR+MS systems which have orbital periods longer than those of low-mass systems.

\begin{figure}
    \centering
    \includegraphics[width=\columnwidth]{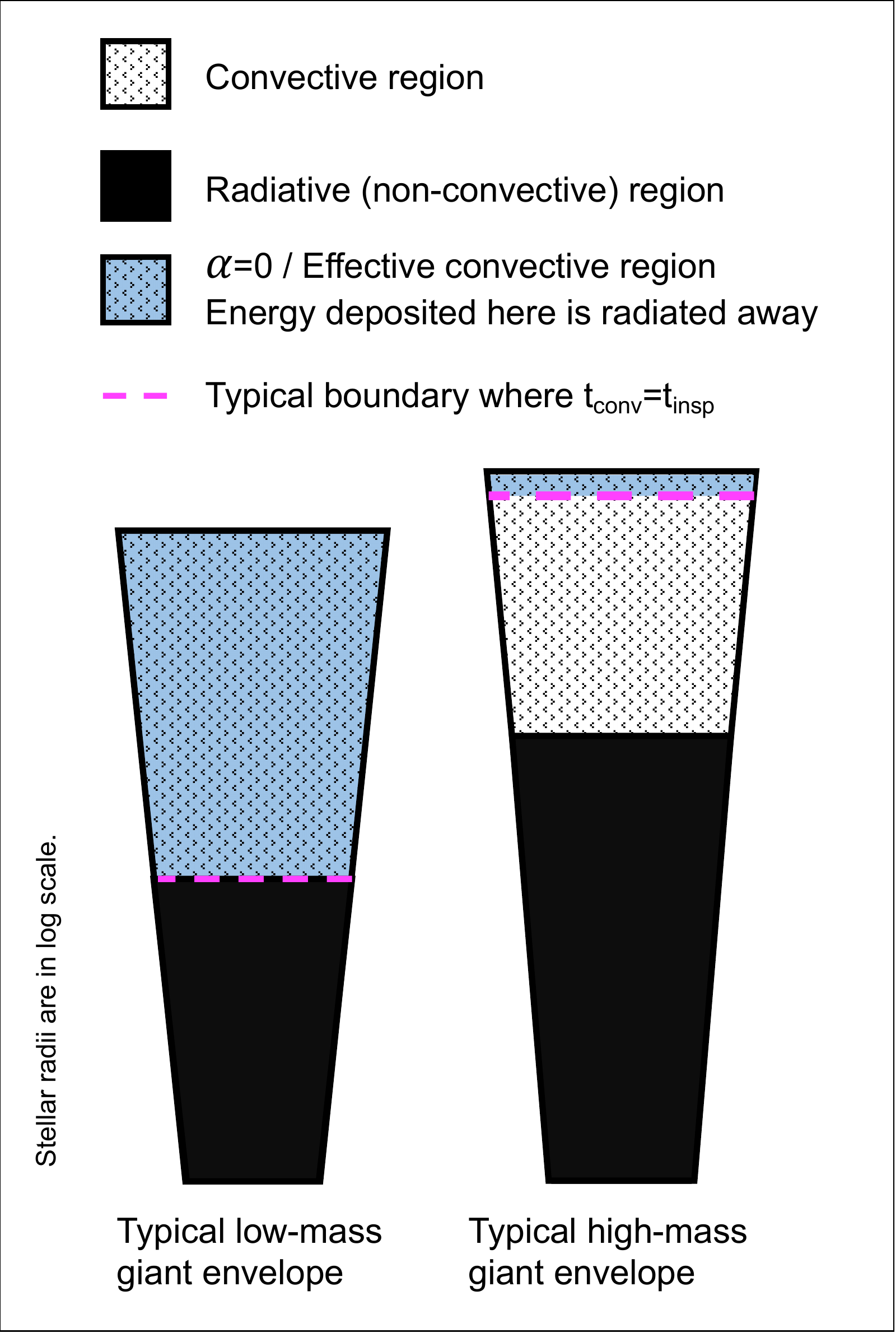}
    \caption{A cartoon showing the difference in low-mass and high-mass envelopes of (super)giants. While the convective regions of both are deep and a substantial part of the envelope, the region where convection is effective at removing the companion's orbital energy is dramatically different between the two. In low-mass stars, the companion can travel much deeper into the envelope before its energy is tapped to drive ejection, thus lowering the ejection efficiency, such that $\alpha \ll 1$; in high-mass stars, the opposite is true and the envelope is ejected closer to the surface, such that $\alpha\approx 1$. }
    \label{fig:cartoonenvelopes}
\end{figure}

\begin{figure}
    \centering
    \includegraphics[width=\columnwidth]{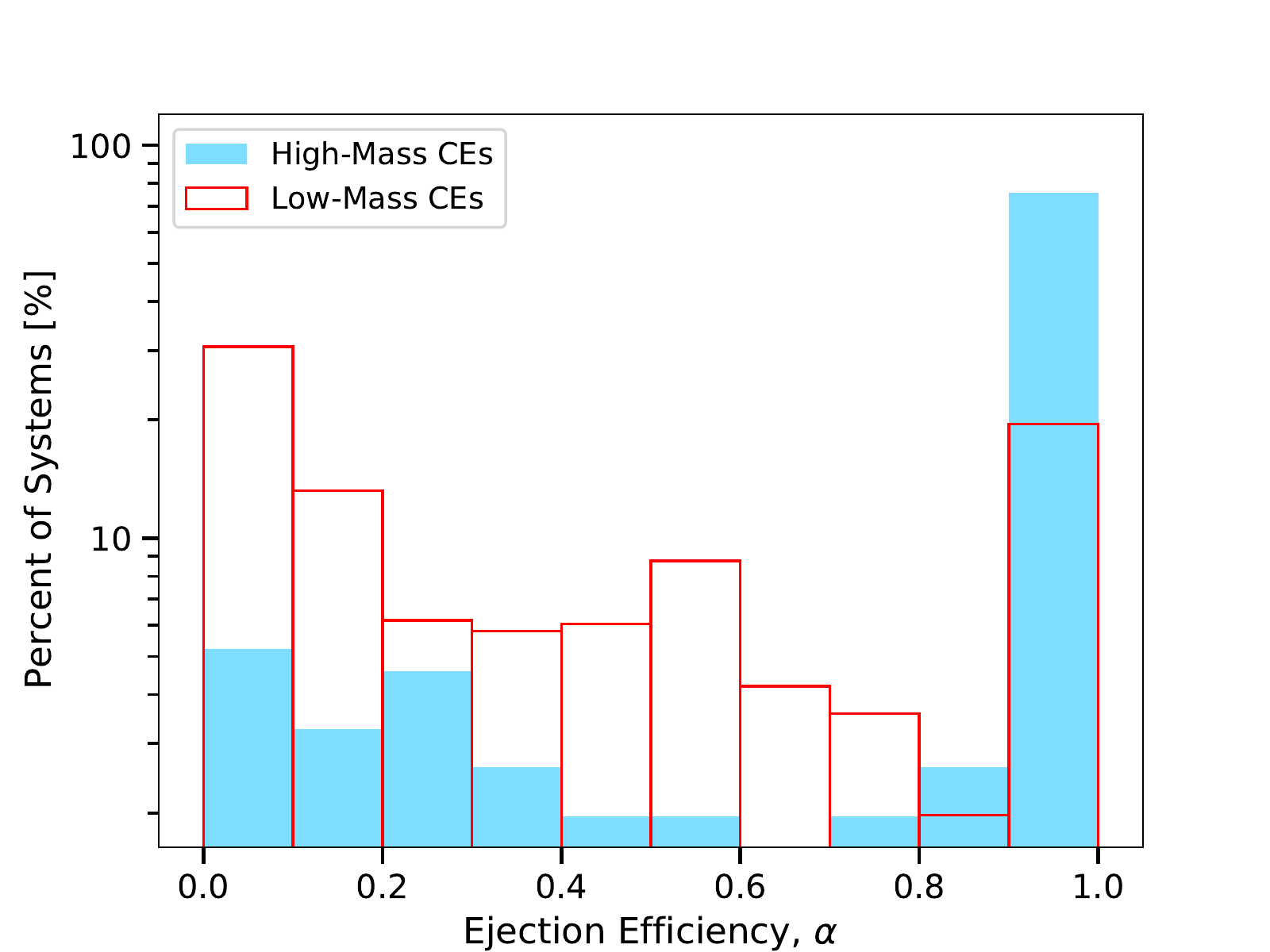}
    \caption{A histogram of CE efficiencies, $\alpha$, for high-mass and low-mass CEs. The y-axis is the percent of systems in the sample such that $y=\int_0^1 N(\alpha) {\rm d}\alpha=100$. Solid blue bars represent efficiencies which are determined alongside the final separations shown in Fig.~\ref{fig:rvst}(b). The vast majority of the systems are produced with high ($\alpha\rightarrow1$) efficiencies. In contrast, the low-mass efficiencies shown in red outline are as calculated by \citet[][Fig. 6]{Wilson2019} and tend to be lower.}
    \label{fig:alphahist}
\end{figure}

Though both low-mass and high-mass giant stars possess deep and vigorous convective envelopes, convection is only effective at transporting energy in cases where $\tconv<\tinsp$.  In low-mass stars, the inspiral times of companions with sufficient energy to unbind the envelope are long compared to the convective transport timescale; in high-mass stars, the inspiral times of companions with sufficient energy to unbind the envelope are short compared to the convective transport timescale. This is summarized in Fig.~\ref{fig:cartoonenvelopes}. For this reason, in high-mass envelopes, the efficiency of the CE interaction, $\alpha$, is high, such that convection within the envelope cannot transport the companion's orbital energy away from the system. Instead, the vast majority of the energy deposited by the companion contributes to ejecting the envelope rather than escaping the system. This high efficiency indicates that the envelope will be ejected, halting the companion's inspiral and therefore establishing a final separation, close to the surface of the star. A histogram of the efficiency values which were determined alongside the final separations of high-mass post-CE binaries are shown in Fig.~\ref{fig:alphahist}. While we do not draw from a functional IMF for mass ratios and initial separations, we do use a set prescription for primary and companion masses modelled. These are described in Sec.~\ref{sec:finalsepshift}. The efficiencies are calculated from the time of engulfment ($a_{\rm init}=R_\star$) for different times in the primary's evolution, as shown in Fig.~\ref{fig:rvst}(a).

The change in orbital energy of the companion, $\Eorb$, is sufficient to unbind a massive star's envelope at a radius which is consistent with observations of WR binaries. The inclusion of convection acts to reduce the energy which can contribute to unbinding the shared envelope (decreases the ejection efficiency). In low-mass stars, convection is effective at carrying energy of the companion to the surface of the star where it is radiated away; in high-mass stars convection is \textit{not} effective at carrying the energy of the companion for two reasons. First, the companions which have sufficient energy to unbind the envelope also have orbital decay timescales which are short. Second, the maximum luminosity able to be accommodated by subsonic convection is exceeded in many cases during inspiral of these higher-mass companions. Unlike low-mass binaries which require being spun-up to only a small fraction of co-rotation, high-mass binaries would require high (or non-existent) values of spin-up for the convection to remain subsonic \citep[see Equation 8 in][]{Wilson2020}. Despite the overall ineffectiveness of convection at reducing the efficiency of the CE interaction in high-mass stars, the radius where $\alpha \Eorb \ge \Ebind$ (in some cases modulated by values described in Sec~\ref{sec:finalsepshift}) continues to match the populations of binaries which are progeny of the initial masses used.

The final separations found in this work, while consistent with observations, do not incorporate the effects of radiation pressure of the massive star's envelope, which may have an effect on the companion's inspiral \citep{Lau2022}. In addition, all primary stars were assumed to have solar metallicity (Z=0.02). Changes in the metallicity of the supergiant may have an effect on the envelope's binding energy and consequent CE outcomes \citep{Aguilera-Dena2021,Klencki2021}.

\subsection{Implications for Population Synthesis Codes and Gravitational Wave Sources}

The efficiency of the CE interaction can be described as a function of the internal structure of the star rather than a constant value. In low-mass CEs, we have shown that $\alpha$ is low in most cases, but that the particular values of $\alpha$ vary dramatically given different convective depths, mass ratios, and stellar phases \citep{Wilson2019}. These structure-dependent efficiency values match observations of M dwarf + white dwarf binaries and double white dwarfs, reinforcing the importance of convection on determining the final separations of CE interactions \citep{Wilson2020}. In high-mass binaries, however $\alpha$, while still dependent on the internal structure of the shared envelope, is much higher, revealing a more efficient CE phase. The final separations which follow from an efficient CE interaction reproduce observations of post-CE WR binaries. Current and future population synthesis codes, such as \texttt{POSYDON} \citep{Fragos2022}, which can provide variable $\alpha$ values for binary scenarios, may better match observations with considerations of the internal structure of the primary. Without detailed internal structure of the binary, a reasonable approximation is to set $\alpha=1$ for CE interactions with high-mass primaries.

CE evolution is the primary pathway by which binary stars migrate to short-period orbits that eventually decay by emission of gravitational waves. Understanding the processes prior to detectable gravitational-wave emission provides a more complete understanding of merging compact objects. For high-mass stars, we provide a physical motivation for and confirm the assumption that the CE interaction is very efficient, meaning that $\alpha$ which best describes the CE phase that produces the progenitors of gravitational-wave sources is high  \citep[$\alpha\approx 1$, e.g.,][]{Belczynski2002,Belczynski2020}. On the other hand, the CE interaction for low-mass stars is very \textit{inefficient}, meaning that $\alpha$ which best describes the CE phase which produces WD+WD binaries is low \citep{Wilson2020}.

\section{Conclusion}
\label{sec:Conclusion}

In this paper, we investigated the effect of convection on the common envelope phase of high-mass stars and determined that the discrepancy in the ejection efficiency of high-mass and low-mass CEs can be explained by including convection in the CE phase. We have shown that common envelope evolution with the inclusion of convection matches observations of envelope-stripped Wolf-Rayet binaries (a high-mass post-CE binary) by comparing to final separations of high-mass binaries which undergo a CE interaction. Using detailed stellar interior models, we consider the effects of convection on the final systems which originate from high-mass binaries. We find the following.

\begin{itemize}
    \item[--] The inclusion of convection in CE evolution provides a way to explain the discrepancy in efficiencies inferred from observations of low-mass and high-mass post-CE binaries.    
    \item[--] The change in orbital energy of the companion is sufficient (no additional energy sources are required) to unbind the envelope. Final separations of post-CE binaries match those of observations when convection is considered. This is true for both high-mass and low-mass CE interactions (see Fig.~\ref{fig:rvst}(b)).
    \item[--] Although high-mass stars have deep and vigorous convective envelopes, convection has little or no impact on the radius at which the envelope becomes unbound. This is largely due to short inspiral timescales. This is in contrast to the large effect that convection has on the CE phase and final separations of low-mass binaries \citep[][see Fig.~\ref{fig:cartoonenvelopes}]{Wilson2019}. 
    \item[--] The efficiency of high-mass CE evolution is high ($\alpha \approx 1$), whereas the efficiency is low in low-mass systems due to the difference in the role of convection in high-mass and low-mass envelopes (see Fig.~\ref{fig:alphahist}).
\end{itemize}

Wolf-Rayet binaries evolve to form different compact binaries with neutron star (NS) and black hole (BH) components. Characterizing the nature of high-mass common envelope evolution is an important step in our understanding of the progenitors of gravitational-wave sources. Other massive-star binaries, such as the Wolf-Rayet stars explored here, act as laboratories to study post-CE outcomes of massive stars.

Future work should incorporate convection and radiation in high-resolution numerical studies of common envelope evolution.  This would allow more robust conclusions to be drawn as to the effect of convection on CE interactions in both low-mass and high-mass binary systems. In addition, a similar analysis as that conducted in this work can be extended more directly to the formation channels of compact-binary gravitational wave progenitors wherein a NS or BH companion is immersed in a convective CE. 

\section*{Acknowledgements}

ECW and JN acknowledge support from the following grant: NSF~AST-2009713. The authors thank Christopher Tout, Maria Drout, Morgan MacLeod, Joel Kastner, Silvia Toonen, Eric Blackman, Adam Frank, and Luke Chamandy for stimulating discussions regarding this work.

\section*{Data Availability}

No new observational data were generated from this research. Data underlying this article are available in the articles and supplementary materials of the referenced papers. Stellar interior models were derived from Modules for Experiments in Stellar Astrophysics (\texttt{MESA}) which is available at http://mesa.sourceforge.net.



\bibliographystyle{mnras}
\bibliography{allbibs} 


\bsp	
\label{lastpage}
\end{document}